\documentclass{article}[12pt]
\usepackage{epsfig}

\def\la{~\mbox{\raisebox{-.6ex}{$\stackrel{<}{\sim}$}}~}
\def\ga{~\mbox{\raisebox{-.6ex}{$\stackrel{>}{\sim}$}}~}

\begin{document}

\begin{center}
{\Large {\bf Completing Natural Inflation}
}
\\[0pt]

\bigskip
\bigskip {
{\bf Jihn E. Kim $^{a,}$\footnote{ {{ {\ {\ {\ E-mail:
jekim@phyp.snu.ac.kr}}}}}}}, {\bf Hans Peter\ Nilles
${}^{b,}$\footnote{ {{ {\ {\ {\ E-mail:
nilles@th.physik.uni-bonn.de}}}}}}}, {\bf Marco\ Peloso
${}^{c,}$\footnote{ {{ {\ {\ {\ E-mail:
peloso@physics.umn.edu}}}}}}}
\bigskip }\\[0pt]
\vspace{0.23cm}
{\it ${}^b$ School of Physics, Seoul National University,}\\
{\it Seoul 151-747, Korea,} \\
\vspace{0.23cm}
{\it ${}^b$ Physikalisches Institut der Universit\"at Bonn,} \\
{\it Nussallee 12, 53115 Bonn, Germany.}\\
\vspace{0.23cm}
{\it ${}^c$ School of Physics and Astronomy,}\\
{\it University of Minnesota, Minneapolis, MN 55455, USA.}\\

\end{center}

\vspace{1cm}

{\bf \large Abstract}

\vspace{0.5cm}

If the inflaton is a pseudo-scalar  axion, the
axion shift symmetry can protect the flatness of its potential
from too large radiative corrections. This possibility, known as
natural inflation, requires an axion scale which is greater than
the (reduced) Planck scale. It is unclear whether such a high
value is compatible with an effective field theoretical
description, and if the global axionic symmetry survives quantum
gravity effects. We propose a mechanism which provides an
effective large axion scale, although the original one is
sub-Planckian. The mechanism is based on the presence of two
axions, with a potential provided by two anomalous gauge groups.
The effective large axion scale is due to an almost exact symmetry
between the couplings of the axions to the anomalous groups. We
also comment on a possible implementation in heterotic string
theory.

\vspace{1cm}

\section{Introduction and Discussion}

There is compelling evidence that the early universe underwent
a period of inflation. Although several different models for
inflation exist, they all share the nontrivial requirement of the
flatness of the inflaton potential. This is mandatory for (i)
inflation itself to take place, and (ii) to match the primordial
perturbations indicated by the CMB anisotropies, namely an almost
scale free spectrum of scalar perturbations, with an amplitude of
the order of $10^{-5} \,$. The requirement (ii) is by itself very
restrictive. For instance, if the inflaton potential can be
approximated by a mass term~\cite{chaos}, $V \sim m^2 \, \phi^2 / 2
\;$, inflation can occur at large $\phi\,$, irrespectively of the
inflaton mass $m\,$. However, the correct spectrum of perturbations
is achieved for $m \sim 10^{13} \, {\rm GeV} \,$, five orders of
magnitude smaller than the most natural ({\it a-priori}) expectation
$m \sim M_p$ (in this paper, $M_p \simeq 2.4 \times 10^{18} \,
{\rm GeV} \,$ denotes the reduced Planck scale).

In particle physics models, the flatness of the potential should
be protected against radiative corrections, which can arise from
the inflaton self-interactions (which are also severely
constrained) or from its coupling to matter fields, responsible
for the reheating of the universe after inflation. As for the
Higgs field in the standard model of particle physics,
supersymmetry can provide a natural protection against radiative
corrections. In some cases, for instance some versions of hybrid
inflation~\cite{hybrid}, supersymmetry can play even a more
crucial role (for a review, see~\cite{tony}): if the inflaton
direction is exactly flat in the limit of unbroken supersymmetry,
supersymmetry breakdown can provide a small (logarithmic) tilt of
the inflaton direction, which allows for a correct spectrum for
the perturbations.

Another possible symmetry which can protect the flatness of the
inflaton potential is an axionic (shift) symmetry, in the case in
which the inflaton is a pseudo-scalar axion. In this
scheme, the inflaton potential arises due to the breaking of a
(global) axionic symmetry $\phi \rightarrow \phi + {\rm
constant}\,$, and it is therefore controlled by it: for instance,
the couplings of the inflaton to matter do not affect the inflaton
potential if they respect the axionic symmetry. This mechanism,
known as natural inflation, was originally proposed
in~\cite{katie}, and several possible implementations have been
discussed in~\cite{bond}. A shift symmetry has for instance been
advocated in supergravity (either by itself or as part of a more
general no scale symmetry~\cite{noscale}) to protect the inflaton
potential against K\"ahler corrections. Shift symmetries also
arise within string theory, and their application to inflation has
been considered for instance in~\cite{bond,large}, and - within
KKLT~\cite{kklt} compactification - in~\cite{kallosh}.

The breaking of the axionic symmetry leads to the inflaton
potential
\begin{equation}
V =   \Lambda^4 \left[ 1 - {\rm cos } \left( \phi / f \right)
 \right] \,\,,
\label{pot}
\end{equation}
where $f$ is the axion scale (entering in $V$ after $\phi$ has
been canonically normalized). Inflation occurs while $\phi$ is
close to a maximum of $V\,$, where the flatness condition
$\epsilon \equiv M_p^2 V'^2 / 2 V^2 \ll 1$ is satisfied.~\footnote{The range
of $\phi$ for which this condition is met shrinks considerably as
$f$ decreases below $M_p \,$. Hence, a sufficient amount of
inflation becomes unlikely at small $f\,$. It is however difficult
to translate this consideration into a strict lower limit for
$f\,$, since a ``measure'' in the space of initial conditions is
not uniquely defined. See~\cite{fk} for a detailed discussion.}
However, the second flatness condition close to the maximum reads
$\vert \eta \vert \equiv \vert M_p^2 V'' / V \vert = M_p / 2 f^2 \ll 1\,$,
setting a direct limit on the axion scale. The limit is obtained by
computing the spectral index for the primordial perturbations, which in the slow roll
approximation~\footnote{A numerical calculation showed that this expression
is accurate as long as $f \la 7.5 \, M_p\,$~\cite{bond}.}  is~
\begin{equation}
n_s = 1 - 6 \, \epsilon + 2 \, \eta \simeq 1 - \frac{M_p^2}{f^2}
\end{equation}
(in the last expression we have assumed $\epsilon \ll \eta \,$, which is appropriate
as long as $\phi$ is close to the maximum). Hence, the spectrum ``reddens'' as
$f$ decreases.  The WMAP limit on the spectral index $\vert n_s - 1 \vert \la 0.1 \,$ (as
computed in~\cite{ns}) translates into the bound~\cite{fk}~\footnote{The
amplitude of the fluctuations ($\delta \rho / \rho \sim 10^{-5}$) does not
set a further direct limit on $f\,$, since it also depends on the scale of the potential $\Lambda\,$.
If the bound~(\ref{bound}) is saturated, the correct amplitude is obtained for
$\Lambda \simeq 10^{15} \, {\rm GeV} \;$~\cite{fk}.}
\begin{equation}
f \ga 3 \, M_p \,\,.
\label{bound}
\end{equation}

It is legitimate to wonder whether such a high value is compatible
with an effective field theory description~\cite{large,paolo}. In
particular, it can be expected that for high $f$ quantum gravity
effects will break the global axionic symmetry~\cite{graglo}. In
that sense, equation~(\ref{bound}) is the main stumbling block for
natural inflation. String theory realizations have further
problems to accomodate a large $f$, as emphasized in~\cite{large}.
For instance, in the simplest version where the inflaton is
associated to the model independent axion of heterotic string
compactifications, the scale $f$ is related to the value of the
dilaton field, and the required value of $f$ is in the strong
coupling regime, where the supergravity description breaks down.

Some versions of inflation, as for instance the
hybrid~\cite{hybrid} or the assisted~\cite{assisted} ones, make a
nontrivial use of two or more scalar fields. As we show in this
paper, the presence of two or more axions can also have
interesting consequences, and it can result in a solution to the
problems mentioned above. More precisely, it is possible to obtain
some directions characerized by an effective axion scale which is
much larger than the ones of the original fields. To illustrate
the general idea, it is sufficient to consider two axionic fields
$\theta \,, \rho$ (the extension to more fields is trivial), with
a potential
\begin{equation}
V = \Lambda_1^4  \left[ 1 - {\rm cos } \left( \frac{\theta}{f_1}
+ \frac{\rho}{g_1} \right) \right] +
\Lambda_2^4  \left[ 1 - {\rm cos } \left( \frac{\theta}{f_2} +
\frac{\rho}{g_2} \right) \right] \,\,
\label{pot2}
\end{equation}
It is easy to see that, when the condition
\begin{equation}
f_1 / g_1 = f_2 / g_2
\label{condition}
\end{equation}
is met, the same linear combination of the two axions (denoted by
$\psi$) appears in both terms of~(\ref{pot2}). Hence, the
orthogonal combination $\xi$ is a flat direction of $V \,$. In
general, the lifting of the potential along $\xi$ is suppressed as
long as the condition~(\ref{condition}) holds at an approximate
level. In this case the field $\xi$ can be a good inflaton
candidate, even when the scales $f_{1,2} \,,\, g_{1,2}$ are all
smaller than $M_p \,$. This is the main result of our paper. It
allows us to circumvent the bound in equation~(\ref{bound}) and
thus removes a severe problem of natural inflation.

The equality~(\ref{condition}) can be
accidental, or due to a symmetry between the mechanisms
responsible for the breaking of the two shift symmetries $\theta
\rightarrow \theta + C \;,\; \rho \rightarrow \rho + C' \,$. In
the second case, the flat direction $\xi$ will be lifted due to
the breaking of this symmetry, so that a small breaking will
ensure that the $\xi$ direction is sufficiently flat.

This mechanism works independently of the values $\Lambda_{1,2}$
of the two terms in the potential. However, it has a
simpler interpretation when one of the two scales is
significantly larger than the other. In this case, the potential
for the field $\xi$ (once $\psi$ has settled to its minimum) also
looks like a typical axion potential~(\ref{pot}), but with an
effective coefficient $f$ much greater than $g_{1,2} \,$. In this
sense, one can say that the condition~(\ref{condition}) is
responsible for a large effective axion scale.

As a theoretical motivation for two axions, we note that many
models have in their spectrum several axionic fields. For example,
heterotic string theory compactified to $4$ dimensions has model
dependent axions in addition to the model independent one. The
axions receive a potential from their coupling to anomalous gauge
groups.
At tree level in the string
coupling, only the model independent axion (partner of the dilaton
field) is coupled (with a universal strength) to the gauge groups.
However, at the loop level also the model dependent axions are
coupled, with a strength dependent on the specific gauge
group~\cite{ibanez}, as well as on the particle content of the
model. Degeneracies among these couplings result in flat
directions of the potential. This opens up the possibility for
natural inflation, which, as we mentioned, is quite problematic if
only one axion is dynamically relevant~\cite{large}.

It is appropriate to compare the mechanism we are discussing with
some other recent proposals. The possibility of an effective axion scale
$\gg M_p$ has also been discussed in~\cite{paolo}. In that case
the axion direction is associated with the Wilson line of a $U
\left( 1 \right)$ field along one extra-dimension compactified on
a circle of radius $R \,$. Due to the Casimir energy of fields
living in compact space~\cite{hosotani}, the
potential for the Wilson line is also of the form~(\ref{pot}),
with an effective axion scale $f \sim 1 / \left( g_4 \, R \right)
\,$, $g_4$ being the coupling constant of the gauge field in the
effective $4$d theory. If $g_4 \ll 1\,$, the effective scale can
be sufficiently large even when the radius $R \gg M_p^{-1} \,$, so
that quantum gravity effects can be neglected.

A second mechanism which shares some analogy with the present one
has been recently discussed in~\cite{jj}. In this model, the
inflaton is a linear combination of the real and imaginary
directions of a modulus $T\,$, with a superpotential
\begin{equation}
W = W_0 + A \, {\rm e}^{-2 \, \pi\,T/N} + B \, {\rm e}^{-2 \,
\pi\,T/M}
\end{equation}
arising from gaugino condensation in the group ${\rm SU } \left( N
\right) \times {\rm SU } \left( M \right) \,$. For arbitrary $N
,\, M \,$, the resulting potential is not flat enough to support
inflation. However, for $N=M \,$ the imaginary component of $T$
becomes an exact flat direction. Hence, the model can accomodate
inflation provided $N \simeq M \,$ (the values $N=100 \;,\; M=90$
were considered in~\cite{jj}).

The remainder of this paper is devided in two Sections.
Section~\ref{details} describes the mechanism and the evolution of
the two fields in more details, while Section~\ref{concl} presents
our conclusions.

\newpage
\section{A large effective axion scale}~\label{details}

Let us start by assuming $f_1 = f_2 \equiv f \,$ in the
potential~(\ref{pot2}). This is what happens for instance in the
heterotic string case mentioned above, where $f$ is the scale of
the model independent axion (coupled with the same strength to
both the gauge groups). Clearly, this assumption is not necessary
for the mechanism to work, but it simplifies the following
algebra. For example, the condition~(\ref{condition}) is replaced
by the simpler $g_1 = g_2 \,$. As in~\cite{jj}, we can assume that
the potential~(\ref{pot2}) arises from two anomalous groups. For a
hidden sector gauge group $SU \left( N \right) \,$, with $n$ pairs
of light hidden sector quark and anti-quarks in the
(anti-)fundamental representation, the coefficient of the
instanton induced potential is
\begin{equation}
\Lambda^4 = m_Q^a \, m_{\tilde G}^b \, L^{4-a-b} \,,
\label{lambda}
\end{equation}
where $a>0, b>0,$ and $m_Q, m_{\tilde G}$ are the hidden sector
quark and gaugino masses, respectively, while $L$ is the
renormalization group invariant scale of the hidden
sector.~\footnote{Hence, the scale $\Lambda$ can be considerably
lower than $L\,$, and it is actually vanishing if supersymmetry is
unbroken.} If this choice is made, the approximate symmetry
responsible for the flatness of the potential is a symmetry
between the couplings of the two axions to the anomalous groups.
However, the mechanism we have in mind is rather general, and it
can take place independently of the origin of the
potential~(\ref{pot2}).

The two axions $\theta$ and $\rho$ are not mass
eigenstates of the system. The two ``physical'' fields, which we
will denote by $\psi, \xi\,$, are instead the eigenvectors of the
mass matrix at each point in the $\left\{ \theta ,
\rho \right\}$ space, and hence they are linear combinations of
the two axions. Expanding the potential about the minimum $
\theta = \rho = 0\,$, the mass matrix has the determinant
\begin{equation}
\vert {\cal M}^2 \vert \equiv \vert V'' \vert = \frac{\left(
g_1 - g_2 \right)^2 \, \Lambda_1^4 \,
\Lambda_2^4}{g_1^2 \, g_2^2 \: f^2} \,\,.
\label{det}
\end{equation}

We see that there is a flat direction for $g_1 =g_2\,$, where the
condition~(\ref{condition}) is met. As we mentioned, this is
easily understood: in this case both terms of the
potential~(\ref{pot2}) are function of the same linear combination
of $\theta$ and $\rho\,$. The orthogonal combination is thus
massless.
In general, the case $g_1 \simeq g_2$ corresponds
to a situation where one of the linear
combination leads to a smaller mass scale than the naive
expectation $m^2 \sim \Lambda^4 / f^2 \,$ would suggest.
If this direction is
associated with the inflaton field, we obtain a suppression of the
inflaton mass, or equivalently of the $\eta$ parameter
proportional to
$V'' / V \,$. This can guarantee a sufficiently flat spectrum for
the scalar perturbations. From eq.~(\ref{det}) we see that the
suppression takes place for any value of $\Lambda_{1,2} \,$. The
case $\Lambda_1 \gg \Lambda_2$ is, however, more transparent and
therefore we will assume it for simplicity. At leading order in
$\left( \Lambda_2 / \Lambda_1 \right)^4 \,$ one of the two
physical fields (say $\psi$) is the linear combination appearing
in the first term of eq.~(\ref{pot2}), while the second physical
field ($\xi$) is the orthogonal combination; more precisely (up to
subdominant $\left( \Lambda_2 / \Lambda_1 \right)^4 \,$
corrections)
\begin{eqnarray}
&& \psi \equiv \frac{f \, g_1}{\sqrt{f^2 + g_1^2}} \, \left(
\frac{\theta}{f} + \frac{\rho}{g_1} \right) \;\;\;,\;\;\;
m_\psi^2 = \left( \frac{1}{f^2} + \frac{1}{g_1^2} \right)\,
\Lambda_1^4 \,\,, \nonumber\\
&& \xi \equiv \frac{f \, g_1}{\sqrt{f^2 + g_1^2}} \, \left( -
\frac{\theta}{g_1} + \frac{\rho}{f} \right) \;\;\;,\;\;\;
m_\xi^2 = \frac{\left( g_1 - g_2 \right)^2}{g_2^2 \left( f^2 + g_1^2
\right)} \, \Lambda_2^4 \,\,.
\label{summary}
\end{eqnarray}
(notice that the product $m_\psi^2 \, m_\xi^2$ agrees with
eq.~(\ref{det})). In terms of the two physical fields, the
potential~(\ref{pot2}) reads
\begin{eqnarray}
V &=& \Lambda_1^4 \left[ 1 - {\rm cos } \left(
\frac{\sqrt{f^2 + g_1^2}}{f \, g_1} \, \psi \right) \right] +
\nonumber\\
&+& \Lambda_2^4 \left[ 1 - {\rm cos } \left(
\frac{f^2 + g_1 \, g_2}{f \, g_2 \, \sqrt{f^2 + g_1^2}} \, \psi +
\frac{g_1 - g_2}{g_2 \, \sqrt{f^2 + g_1^2}} \, \xi \right)
\right] \,\,.
\label{potphys}
\end{eqnarray}

We explicitely see that, for $g_1 = g_2\,$, only the combination
$\psi$ is coupled to the anomalies, and this explains why $m_\xi^2
\propto \left( g_1 - g_2 \right)^2 \,$ in the previous equation.
The potential in terms of the physical fields $\psi$ and $\xi$ has
the characteristic form of an axion potential. For this reason we
can interpret the suppression of the mass of $\xi$ as an increase
of the corresponding effective axion scale $f_\xi \,$,
\begin{equation}
f_\xi \equiv \frac{g_2 \: \sqrt{f^2 + g_1^2}}{\vert g_1 -
g_2 \vert} \,\,.
\label{effscale}
\end{equation}

\begin{figure}
\centerline{\includegraphics[width=0.6\hsize]{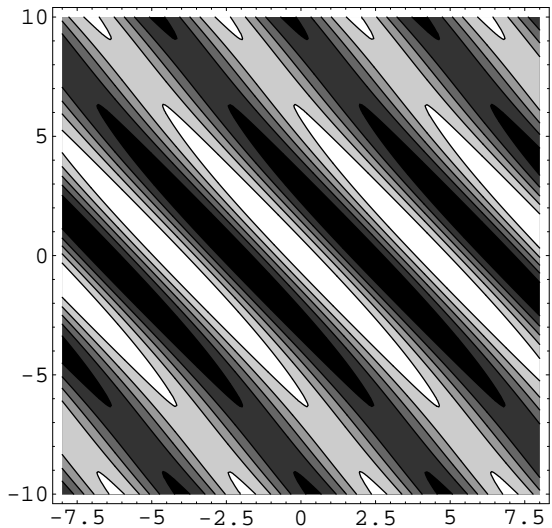}}
\caption{Contour plot of the potential, for $\Lambda_1 = 1.5 \,
\Lambda_2\,$, $f=g_2 = 0.7 \, M_p \,$, $g_1 = 0.98 \, M_p \,$
(giving $f_\xi \simeq 3 \, M_p \,$). See the main text for details.}
\label{fig1}
\end{figure}

For illustrative purposes, a contour plot of the potential is
shown in fig.~\ref{fig1}.
Black regions corresponds to high $V\,$, while
white ones to low values. The horizontal axis is $\theta \,$,
while the vertical one is $\rho \,$, both given in units of
$M_p \,$. The inflaton $\xi$ is almost aligned along the diagonal
from up left to down right, while the heavier field $\psi$ is
essentially aligned along the other diagonal. The difference between the
scales of $\psi$ and $\xi$ is manifest in the figure.

The cosmological evolution of the two fields is the following: due
to the difference in the scales of the two terms, the two axions
$\psi$ and $\xi$ evolve independently; first, $\psi$ evolves under
the effect of the first term (the second term giving a negligible
contribution), oscillating around the minimum $\psi = 0\,$. The
second axion remains
essentially frozen at its initial value during this
stage. If this value is close to a maximum of $V \left( \xi \right)$
(more accurately, to a saddle point
of the overall potential, since $\psi$ is in its minimum), this term will
eventually dominate, and $\xi$ will then drive inflation. We
identify this second stage with the observed stage of inflation.
The mass of $\psi$ is much higher than $H$ during this stage, and
$\psi$ quickly settles to the minimum $\psi = 0$. Hence, $\psi$ is
dynamically irrelevant during the observable stage, and $\xi$
simply evolves under the effect of the second term in $V\,$.

\newpage
\section{Conclusions}~\label{concl}

We considered  natural inflation with two axions (with two decay
constants $f_1$ and $f_2$) and  two confining gauge groups.
This allows us to circumvent a serious problem of natural
inflation coming from equation~(\ref{bound}) and the potential importance
of quantum gravitational effects.
We conclude our discussion by explaining that
the present mechanism is not
destroyed  by such quantum gravitational effects
as long as $f_i < M_P$ for all $i$.

The anomalous couplings to two nonabelian groups can be different,
with effective decay constants  $f_1\epsilon_1,
f_1\epsilon_2, f_2\epsilon_3,$ and $f_2\epsilon_4$, where
\begin{eqnarray}
{\cal L}_{\rm axion\
coupling} &=& \frac{a_1}{f_1}\left(\frac{(1/\epsilon_1)}{32\pi^2}F_1\tilde
F_1+ \frac{(1/\epsilon_2)}{32\pi^2}F_2\tilde F_2\right) + \nonumber\\
&+& \frac{a_2}{f_2}\left(\frac{(1/\epsilon_3)}{32\pi^2}F_1\tilde F_1+
\frac{(1/\epsilon_4)}{32\pi^2}F_2\tilde F_2\right)
\end{eqnarray}
and $F\tilde F=\frac12\epsilon^{\mu\nu\rho\sigma}
F_{\mu\nu}F_{\rho\sigma}$. Note, however, that $(1/\epsilon_i)$
are just the expressions for the axion couplings to the anomaly
(which are e.g. determined by the axial charges of fermions),
while the decay constants corresponding to the Goldstone bosons
are simply $f_1$ and $f_2$. Thus, the axionic couplings to matter
are determined by $f_1$ and $f_2$, namely $\sim
(1/f_i)(\partial^\mu a_i) J_\mu^{{\rm (matter)} i}$. For the
gravitational effects, the decay constants are appearing in the
form $f_i/M_P$, and hence for $f_i\ll M_P$ quantum gravitational
effects are negligible in our scenario. The effective large decay
constant (\ref{effscale}) resulting when $g_1\simeq g_2$   is
simply a realization of an almost flat direction in this framework
with sub-Planckian decay constants. Thus, quantum gravitational
effects will not change the flatness of the $\xi$ potential (even
if the effective scale $f_\xi>M_P$) as long as $f_i\ll M_P$, and natural inflation can
be implemented without problems.

\section{Acknowledgments}

It is a pleasure to thank Kiwoon Choi, Katherine Freese, Josh Frieman, Stefan
Groot Nibbelink and Erich Poppitz for useful discussions. The work of
JEK was partially supported by the KOSEF Sundo Grant, the ABRL
Grant No. R14-2003-012-01001-0, and the BK21 program of Ministry
of Education, Korea, and that of HPN by the European Commission
RTN programs \mbox{HPRN-CT-2000-00131}, 00148 and 00152. JEK and
HPN acknowledge the support of the Aspen Center for Physics where
the present work has been completed.

\end{document}